# A Novel Two-stage Entropy-based Robust Cooperative Spectrum Sensing Scheme with Two-bit Decision in Cognitive Radio

**Nan Zhao**
School of Information and Communication Engineering, Dalian University of Technology
Dalian, Liaoning, 116024, China
[zhaonan@dlut.edu.cn]

## *Abstract*

Spectrum sensing is a key problem in cognitive radio. However, traditional detectors become ineffective when noise uncertainty is severe. It is shown that the entropy of Gauss white noise is constant in the frequency domain, and a robust detector based on the entropy of spectrum amplitude was proposed. In this paper a novel detector is proposed based on the entropy of spectrum power density, and its performance is better than the previous scheme with less computational complexity. Furthermore, to improve the reliability of the detection, a two-stage entropy-based cooperative spectrum sensing scheme using two-bit decision is proposed, and simulation results show its superior performance with relatively low computational complexity.

The research is supported by the Fundamental Research Funds for the Central Universities.

## 1. Introduction

**O**ver the last decade, because of conflicts between the increasing demands of wireless communication services and the scarcity of wireless spectrum, cognitive radio (CR) network-related research has progressed rapidly [1]. In CR, the secondary users need to opportunistically sense the idle channels. Once an idle channel is sensed, the secondary users will access the channel. Hence, spectrum sensing requests the secondary users to efficiently and effectively detect the presence of the primary signals, and is a fundamental problem in CR [2]. Generally, spectrum sensing techniques can be classified into three categories, energy detection [3-4], matched filtering detection [5] and feature detection [6]. In the matched filtering detection and feature detection, the CRs should have some knowledge about the primary signal features, such as preambles, pilots, synchronization symbols and modulation schemes. Hence, these two detection schemes require large computational costs and are not suitable to act as a blind detector. Energy detection is shown to be optimal if the cognitive devices do not have a priori information about the features of the primary signals, and it possesses the lowest computational costs and is easily implemented. However, it is susceptible to noise uncertainty and performs poorly at low SNR.

Because of the fluctuation of background noise, noise uncertainty exists in every practical system. Sensitivity to noise uncertainty is a fundamental limitation of current spectrum sensing strategies in detecting the presence of the primary users in CR. Because of the noise uncertainty, the performance of traditional detectors deteriorates quickly when SNR is low [7]. The information entropy theory has been applied to the signal detection successfully, and thus several entropy-based detectors have been proposed to solve the spectrum sensing problem in CR [8-9]. In [8], an entropy-based spectrum sensing scheme is designed by combining the entropy detection in the time domain and the matched filter. However the matched filter in the scheme needs some necessary knowledge about the primary signal features, which requires additional overhead and even hardly holds in CR, and thus it is not a blind detector. In [9], an entropy-based spectrum sensing scheme in the frequency domain based on the spectrum amplitude is proposed and proved to be robust to the noise uncertainty, however, its performance still can be improved.

In order to enhance sensing performance, more sensing time is needed. However, during the process of sensing, secondary users should stop data transmission to avoid being recognized as primary users. Therefore more sensing time means lower secondary system capacity, making this approach less attractive. Cooperative spectrum sensing (CSS) [10-14], where local sensors sense and then send information to the centre where the final decision is made, has been studied extensively as a promising alternative to improve sensing performance. There are mainly three schemes of CSS: AND-rule-based CSS [10], OR-rule-based CSS [11], and VOTING-rule-based CSS [12]. However, these three CSS schemes are rather simple, and their performance is limited. These days, the CSS schemes based on weight have been proposed [13-14] with excellent performance, however, in these schemes SNR of each secondary user should be estimated perfectly to get the fusion weight, and it is difficult to realize.

In this paper, a novel entropy-based spectrum sensing scheme in the frequency domain based on the spectrum power density is proposed, and we prove that it is also robust to the noise uncertainty with better probability of detection and lower computational complexity. To further improve the reliability of the detection, a novel two-stage entropy-based spectrum sensing scheme is designed, which has better performance than those one-stage ones with relatively low computational complexity. Furthermore, a cooperative spectrum sensing scheme with two-bit decision, which is obtained from the two-stage entropy-based sensing results, is proposed. The proposed two-stage entropy-based robust cooperative spectrum scheme can achieve much better performance than AND, OR, and VOTING rule CSS schemes with less computational complexity.

The rest of this paper is organized as follows. In Section II, we describe the system model for spectrum sensing, and the previous entropy detector based on spectrum amplitude is described. In Section III, the novel entropy detector based on spectrum power density is introduced, and its robustness to the noise uncertainty is proved. Two-stage entropy detection scheme is also proposed in Section III. In Section IV, the cooperative spectrum sensing scheme based on two-bit decision getting from two–stage sensing results is proposed. In Section V, the advantages of the proposed cooperative spectrum sensing scheme is illustrated through plenty of simulations. Conclusions are drawn in Section VI.

## 2. System Model

In order to avoid interfering with the primary users when the frequency bands are already occupied, detection should be made before the CR accesses the bands. So the most critical technique is the spectrum sensing which decides success or failure of the following steps. The target of spectrum sensing in CR is to determine whether a licensed band is currently occupied by its primary user or not. This can be formulated into a binary hypotheses testing problem [15]

$$x(n) = \begin{cases} w(n), & H_0 \\ s(n) + w(n), & H_1 \end{cases} \quad (1)$$

where $n=0, 1, …, N$; $N$ is the number of samples. The primary user's signal, the noise and the received signal are denoted by $s(n)$, $w(n)$ and $x(n)$ respectively. $H_0$ represents the absence of primary signal, while $H_1$ represents the presence of primary signal. The noise $w(n)$ is assumed to be additive white Gaussian noise (AWGN) with zero mean and variance of $\sigma_0^2$, and the signal $s(n)$ can either be a deterministic signal (accounting for AWGN channel) or a stochastic signal (corresponding to channel characteristics like fading and multipath) with mean $\mu_1$ and variance $\sigma_s^2$.

Applying discrete Fourier Transform (DFT) to (1), we have the following hypotheses

$$\begin{aligned} H_0 &: \vec{X}(k) = \vec{W}(k), & k=0, 1, ..., K\text{-}1 \\ H_1 &: \vec{X}(k) = \vec{S}(k) + \vec{W}(k), & k=0, 1, ..., K\text{-}1 \end{aligned} \quad (2)$$

where $K$ is the length of DFT equal to sample size $N$, $\vec{X}(k)$, $\vec{S}(k)$ and $\vec{W}(k)$ are the complex spectrum of the receiver signal, primary signal and noise, respectively.

$$\vec{X}(k) = X_r(k) + jX_i(k) = \frac{1}{N}\sum_{n=0}^{N-1} x(n)\exp(-j\frac{2\pi}{N}kn) \quad (3)$$

where $X_r(k)$ and $X_i(k)$ represent the real part and the imaginary part of $X(k)$, respectively. In [9],

an entropy-based spectrum sensing based on spectrum amplitude $X(k) = \sqrt{X_r^2(k) + X_i^2(k)}$ is proposed.

Information entropy is a measure of the uncertainty associated with a random variable. It quantifies information contained in a message and can be written as

$$H(Y) = -\sum_{i=1}^{L} p_i \log_b p_i, \tag{4}$$

where $b$ is the base of the logarithm. In this letter, we define $b$ equal to e. $p_i$ denotes the discrete probability mass function of $Y$. $L$ is the dimension of the probability space.

There are several techniques that can estimate the entropy of a continuous random variable based on a finite number of observations. To reduce the computational complex, we use the simplest approach, histogram-based estimation of the density function [16]. The number of states of the random variable is then equal to the bin number $L$ (dimension of the probability space). Let $k_i$ denote the total number of occurrences in the $i$th bin with $\sum_{i=1}^{L} k_i = N$. The probability in each state $p_i$ is the frequency of occurrences in the $i$th bin, that is, $p_i = k_i/N$. The bin width $\Delta$ can be expressed as

$$\Delta = \frac{Y_{\max} - Y_{\min}}{L}, \tag{5}$$

where $Y_{\max}$ and $Y_{\min}$ represent the maximum and minimum value of random variable $Y$, respectively. Once bin number $L$ is fixed, bin width $\Delta$ varies with the range of the spectrum amplitude.

Replay $p_i$ in (4) by $p_i = k_i/N$, the equation (4) can be rewritten as

$$H(Y) = -\sum_{i=1}^{L} \frac{k_i}{N} \log \frac{k_i}{N}. \tag{6}$$

The entropy detection in [9] makes decision on the information entropy of the spectrum amplitude $X(k) = \sqrt{X_r^2(k) + X_i^2(k)}$ following (6). In hypothesis $H_0$, $X(k) = \sqrt{W_r^2(k) + W_i^2(k)}$ follows Rayleigh distribution, and the information entropy of $X(k)$ can be expressed as

$$H_L(X) = \log \frac{L}{C_1 \sqrt{2}} + \gamma/2 + 1, \tag{7}$$

where $\gamma$ is the Euler-Mascheroni constant, and $C_1 = \sqrt{-2\log(1-\rho)}$. $\rho$ is a large cumulative distribution function (CDF) probability (e.g. 0.99<$\rho$<1), which is calculated from the approximate maximum of variable $X(k)$.

In hypothesis $H_1$, the received signal consists of both primary signal and background noise, and the entropy of spectrum amplitude when $H_1$ is much smaller than that when $H_0$. Hence, the gap of estimated information entropy between $H_0$ and $H_1$ can be utilized to detect the presence/absence of the primary signal. The decision rule is given as

$$H_L(X) = -\sum_{i=1}^{L} \frac{k_i}{N} \log_b \frac{k_i}{N} \begin{cases} \leq \lambda : \text{decide } H_1 \\ > \lambda : \text{decide } H_0 \end{cases} \quad (8)$$

where $\lambda$ is the threshold determined by the false alarm probability (Pf).

In [9], it is also proved that with probability space partitioned into fixed dimensions, the information entropy of the white Gaussian noise (WGN) is a constant, and the entropy detection based on spectrum amplitude is thus intrinsically robust against noise uncertainty.

## 3. Two-stage Entropy Detection Based on Spectrum Power Density

### 3.1 Entropy Detection Based on Spectrum Power Density

The entropy detection based on spectrum amplitude has relatively high performance and is robust to the noise uncertainty, however, the detection performance still can be improved. In this paper, a novel entropy detector based on spectrum power density is proposed, with better performance and lower computational complex.

In hypothesis $H_0$, the received signal is WGN with zero mean and variance $\sigma_0^2$, and the DFT of the received signal can be expressed as

$$\vec{W}(k) = W_r(k) + jW_i(k) = \frac{1}{N}\sum_{n=0}^{N-1} w(n)\exp(-j\frac{2\pi}{N}kn) \quad (9)$$

where $W_r(k)$ and $W_i(k)$ are the real and imaginary part of $\vec{W}(k)$, respectively. $W_r(k)$ and $W_r(k)$ both follow Gaussian distribution with variance of $\sigma_0^2/2N$. So the spectrum power density of the received signal in $H_0$ $W^2(k)=W_r^2(k)+W_i^2(k)$ follows exponential distribution with parameter $\sigma_1^2=\sigma_0^2/2N$. The probability density function (PDF), CDF and differential entropy of $Y(k)=W^2(k)$ can be given as

$$f(Y) = \begin{cases} \frac{1}{2\sigma_1^2} e^{-\frac{Y}{2\sigma_1^2}}, & W \geq 0 \\ 0, & W < 0 \end{cases} \quad (10)$$

$$F(Y) = \begin{cases} 1-e^{-\frac{Y}{2\sigma_1^2}}, & W \geq 0 \\ 0, & W < 0 \end{cases} \quad (11)$$

$$\begin{aligned} h(Y) &= -\int_{-\infty}^{+\infty} f(Y)\log(f(Y))dY \\ &= 1 + \log(2\sigma_1^2) \end{aligned} \quad (12)$$

Therefore, in this letter, we proposed a novel entropy detection scheme based on the spectrum power density $X^2(k)$, and the advantages of this scheme are listed below:

(1) The detection performance is better than the entropy detection in [9], which will be proved in Section V through simulations.

(2) The computational complex of this novel detection scheme is lower than the detection

scheme in [9], which needs additional $N$ square-root operations.

(3) It is also robust to the noise uncertainty with fixed bin number $L$, proved in *Proposition 1*.

*Proposition 1:* With fixed bin number $L$, the information entropy of spectrum power density of the WGN is a constant, hence entropy detection based on spectrum power density is intrinsically robust against noise uncertainty.

*Proof:* Theoretically, the maximum value of an exponential distributed random variable is infinite. We assume that $N$ samples of the exponential distributed random variable $Y=W^2$ has range $[0, Y_{max}]$. Under CDF probability $\rho \to 1$ (e.g. $0.99 < \rho < 1$), $Y_{max}$ can be obtained from (11) as

$$Y_{max} = -2\sigma_1^2 \log(1-\rho) = C_2 \sigma_1^2, \qquad (13)$$

where $C_2 = -2\log(1-\rho)$. Let $L$ represent the number of bins, the bin width $\Delta$ can be written as

$$\Delta = \frac{Y_{max}}{L} = \frac{C_2 \sigma_1^2}{L}, \text{ as } \rho \to 1. \qquad (14)$$

The relationship between information entropy $H(Y)$ and differential entropy $h(Y)$ can be approximately expressed as

$$H(Y) \approx h(Y) - \log \Delta. \qquad (15)$$

Therefore the information entropy of $Y$ can be expressed as

$$\begin{aligned} H_L(Y) &\approx \sum_{i=1}^{L} p_i \log p_i = h(Y) - \log \Delta \\ &= 1 + \log(2\sigma_1^2) - \log \Delta \\ &= 1 + \log(2\sigma_1^2) - \log \frac{C_2 \sigma_1^2}{L} \\ &= 1 + \log\left(\frac{2L}{C_2}\right). \end{aligned} \qquad (16)$$

From (16) we can see that, the entropy based on the spectrum power density of WGN is a constant for a given number of $L$. So the novel entropy detection based on the spectrum power density is robust against the noise uncertainty. The decision rule of this detection scheme can be obtained as

$$H_L(X^2) = -\sum_{i=1}^{L} \frac{k_i}{N} \log_b \frac{k_i}{N} \begin{cases} \leq \lambda : \text{decide } H_1 \\ > \lambda : \text{decide } H_0 \end{cases} \qquad (17)$$

where $\lambda$ is the threshold determined by Pf, and it is set according to the method in [17].

### 3.2 Two-stage Entropy Detection

When SNR becomes lower, the gap between the information entropies calculated in the hypothesis $H_0$ and hypothesis $H_1$ respectively are getting smaller, hence the detection results become unreliable. To improve the performance of the entropy detection based on spectrum power density at low SNR, a two-stage entropy detection scheme is proposed.

The two-stage entropy detection scheme can be represented by the following steps:

*Step 1:* Apply $N$-point DFT to the received signal $x(n)$, and obtain the variable of spectrum power density $Y(k)=X_r^2(k)+X_i^2(k)$, where $k=1, 2, \ldots, N$.

*Step 2:* Calculate the entropy of $Y(k)$ following (17), and we can get the entropy value $H_{L1}(Y)$. The threshold is denoted as $\lambda$, and the decision of the first stage detection can be made following

$$H_{L1}(Y) \begin{cases} \leq \lambda - \Delta_0 : \text{decide } H_1, \text{ and go to } step\ 5 \\ > \lambda + \Delta_0 : \text{decide } H_0, \text{ and go to } step\ 5 \\ \text{else, go to } step\ 3 \text{ to perform the sencond-stage detection} \end{cases} \quad (18)$$

where $\Delta_0$ is a positive parameter, and threshold $\lambda$ can be obtained through a larger number of simulations in hypothesis $H_0$ for a given Pf using this two-stage entropy detection scheme following the method in [17] beforehand. If $H_{L1}(Y)$ is not in $(\lambda-\Delta_0, \lambda+\Delta_0)$, the finial decision is made and jump to *Step 5*; otherwise, the second stage detection will be performed and jump to *Step 3*.

*Step 3:* Apply $N$-point DFT to the received signal $x(n)$ again, and obtain another $Y(k)$.

*Step 4:* Calculate the entropy of $Y(k)$ following (17), and we can get the entropy value $H_{L2}(Y)$ of the second stage. The final decision can be made following

$$\frac{H_{L1}(Y)+H_{L2}(Y)}{2} \begin{cases} \leq \lambda : \text{decide } H_1 \\ > \lambda : \text{decide } H_0 \end{cases} \quad (19)$$

In (19), the decision is finally made by considering the entropy values of these two stages $H_{L1}(Y)$ and $H_{L2}(Y)$, and the detection performance is improved.

*Step 5:* Current detection ends.

In the two-stage scheme described above, if the entropy in the first stage $H_{L1}(Y)$ is out of $(\lambda-\Delta_0, \lambda+\Delta_0)$, the solution is located in the undoubted region. The final decision is made immediately, and the decision is quite accurate. This situation is equal to the one-stage detection with $N$ samples. If $H_{L1}(Y)$ is in $(\lambda-\Delta_0, \lambda+\Delta_0)$, the solution is located in the doubted region, and a second stage detection will be performed. The final decision is based upon both $H_{L1}(Y)$ and $H_{L2}(Y)$, and it is the mean of the detection results of last two $N$-points one-stage detection, using $2N$ samples. Therefore, the performance of two-stage at this situation is close to the one-stage detection using $2N$ points, and the decision is more reliable. Therefore, the two-stage detection scheme can greatly improve the detection performance and make the decision more reliable.

## 4. Cooperative Spectrum Sensing Based on Two-stage Detection

### 4.1 Traditional Cooperative Spectrum Sensing Schemes

We consider a CR network composed of *K* CRs (secondary users) and a common receiver, as shown in **Fig. 1**. We assume that each CR performs spectrum sensing independently and then the local decisions are sent to the common receiver which can fuse all available decision information to infer the absence or presence of the PU.

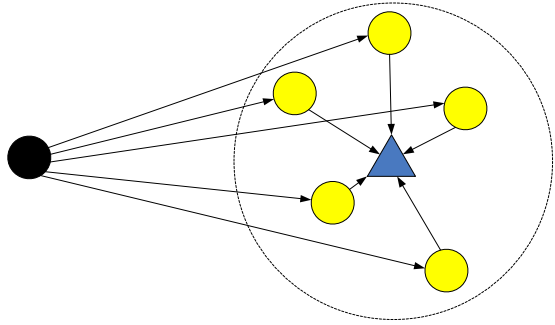

**Fig. 1**. Spectrum sensing structure in a cognitive radio network

In traditional "*n*-out-of-*K*" rule cooperative spectrum sensing, each cooperative partner makes a binary decision based on its local observation and then forwards one bit of the decision $D_i$ (1 standing for the presence of the PU, 0 for the absence of the PU) to the common receiver through an error-free channel. At the common receiver, all 1-bit decisions are fused together according to the logic rule

$$Y = \sum_{i=1}^{K} D_i \begin{cases} \geq n, & H_1 \\ < n, & H_0 \end{cases} \tag{20}$$

where $H_0$ and $H_1$ denote the decision made by the common receiver that the PU signal is *not* transmitted or transmitted, respectively. The threshold *n* is an integer, representing the "*n*-out-of-*K*" rule. It can be seen that the OR rule corresponds to the case of *n*=1, AND rule corresponds to the case of *n*=*K*, and in the VOTING rule *n* is equal to the minimal integer larger than *K*/2.

### 4.2 Cooperative Spectrum Sensing Based on Two-stage Detection

To adapt to the two-stage entropy-based spectrum sensing scheme and further improve the sensing performance, a novel cooperative spectrum sensing scheme is proposed. In the proposed cooperative spectrum sensing scheme, the decision information made by each secondary user includes two bits, which can be calculated from the two-stage detection results.

The proposed cooperative spectrum sensing scheme can be represented by the following steps:

*Step 1:* At the *i*th secondary user, apply *N*-point DFT to the received signal *x*(*n*), and obtain the variable of spectrum power density $Y(k)=X_r^2(k)+X_i^2(k)$, where *k*=1, 2, …, *N*.

*Step 2:* Calculate the entropy of *Y*(*k*) following (17), and we can get the entropy value $H_{L1}(Y)$. The threshold is denoted as *λ*, and the decision of the first stage detection can be made following

$$H_{L1}(Y) \begin{cases} \leq \lambda - \Delta_0 : \text{the final decision } D_i \text{ is set to 11, and go to } step\ 5 \\ > \lambda + \Delta_0 : \text{the final decision } D_i \text{ is set to 00, and go to } step\ 5 \\ \text{else, go to } step\ 3 \text{ to perform the sencond-stage detection} \end{cases} \quad (21)$$

where $\Delta_0$ is a positive parameter, and threshold $\lambda$ can be obtained through a larger number of simulations in hypothesis $H_0$ for a given Pf using this two-stage entropy detection scheme following the method in [17] beforehand. If $H_{L1}(Y)$ is not in ($\lambda$-$\Delta_0$, $\lambda$+$\Delta_0$), the two-bit finial decision of the $i$th secondary user $D_i$ is made and jump to *Step 5*; otherwise, the second stage detection will be performed and jump to *Step 3*.

*Step 3:* Apply $N$-point DFT to the received signal $x(n)$ again, and obtain another $Y(k)$.

*Step 4:* Calculate the entropy of $Y(k)$ following (17), and we can get the entropy value $H_{L2}(Y)$ of the second stage. The final decision can be made following

$$H_{L2}(Y) \begin{cases} \leq \lambda - \Delta_0 : \text{the final decision } D_i \text{ is set to 11, and go to } step\ 5 \\ > \lambda + \Delta_0 : \text{the final decision } D_i \text{ is set to 00, and go to } step\ 5 \\ \text{else, } \dfrac{H_{L1}(Y) + H_{L2}(Y)}{2} \begin{cases} \leq \lambda : \text{the final decision } D_i \text{ is set to 10} \\ > \lambda : \text{the final decision } D_i \text{ is set to 01} \end{cases} \end{cases} \quad (22)$$

The two-bit decision of the $i$th secondary user is finally made by (21) and (22), and the decisions of all the secondary users are then sent to the common receiver to be fused.

*Step 5:* The two-bit decision $D_i$ of each secondary user is received at the common receiver. To transmit it easily and save the spectrum resource, the $D_i$ is two-bit binary, which can be 11, 10, 01, and 00. At the common receiver, to fuse the decisions of all the secondary users together, $D_i$ should be changed into signed integer $F_i$ according to

$$F_i = \begin{cases} 2, & \text{when } D_i = 11; \\ 1, & \text{when } D_i = 10; \\ -1, & \text{when } D_i = 01; \\ -2, & \text{when } D_i = 00. \end{cases} \quad (23)$$

Therefore, we can obtain the final fused decision according to the decisions of all the secondary users in (23) as follows.

$$Z = \sum_{i=1}^{K} F_i \begin{cases} > 0, & H_1 \\ < 0, & H_0 \\ = 0 \begin{cases} H_1, \text{ the number of positive } F_i \text{ is larger than } K/2 \\ H_0, \text{ else} \end{cases} \end{cases} \quad (24)$$

*Step 6:* Current detection ends.

## 5. Simulation Results and Discussion

To evaluate the detection performance of the proposed detection scheme, plenty of simulations are carried out. The signal of the primary user is BPSK modulated, and the baseband symbol rate $f_b$ is equal to 1Mbps. The sampling frequency $f_s$ at the cognitive receiver is 64MHz. The bin number $L$ of the probability space is 15. In all the entropy based detectors, the sample size of DFT is equal to 1024 points. In energy detection, the sample size is also equal to 1024 points.

First, the detection probability (Pd) of these detectors is compared in **Fig. 2** with the power of background noise fixed and Pf=0.1, and the receiver operation characteristic (ROC) curves of these detectors when SNR=-10dB are depicted in **Fig. 3**. The threshold $\lambda$ of the two-stage entropy detection is 1.596, 1.615, 1.629, 1.630 and 1.628 with $\Delta_0$ set to 0.05, 0.1, 0.2, 0.3 and 0.4, respectively, when Pf=0.1. In the simulations, the primary signal experiences Rayleigh fading. From **Fig. 2** and **Fig. 3**, we can see that, the detection performance of the entropy detection based on spectrum power density is better than that of the entropy detection based on spectrum amplitude. The detection performance of the energy detection is a little better than the one-stage entropy detectors. The two-stage entropy detection has the best performance, and the performance becomes better when $\Delta_0$ is larger. When $\Delta_0 \geq 0.2$, the performance remains almost unchanged. In addition, we can see that the detection performance of energy detection is better than two-stage detection ($\Delta_0 \geq 0.2$) when SNR is lower than -13dB and Pd smaller than 0.45. However, we don't care much about the detection performance when Pd is smaller than 0.5, for in this case, the detection probability is relatively small, and missed detection is very easy to happen.

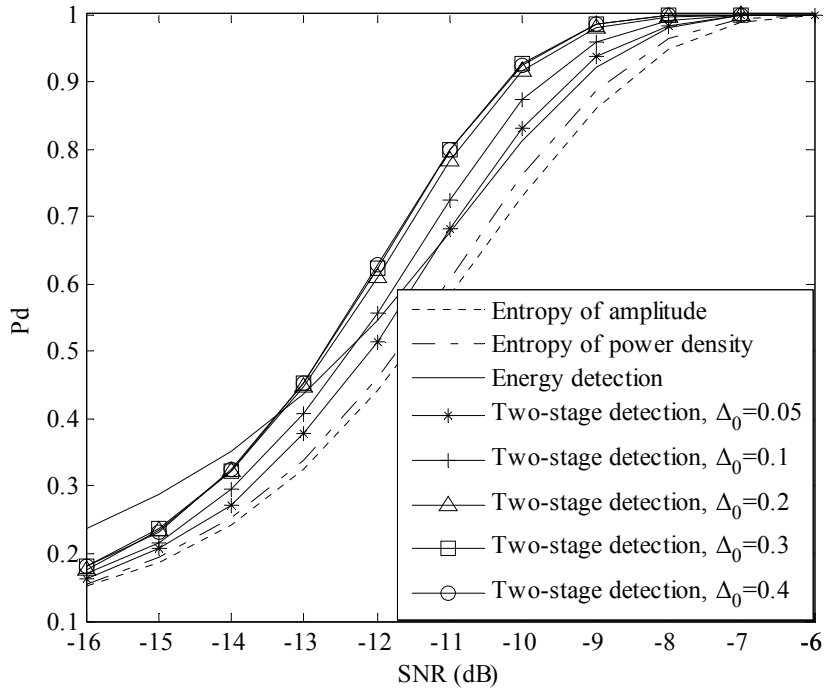

**Fig. 2**. Detection Performance against SNR comparison of the detectors in Rayleigh fading channel without noise uncertainty

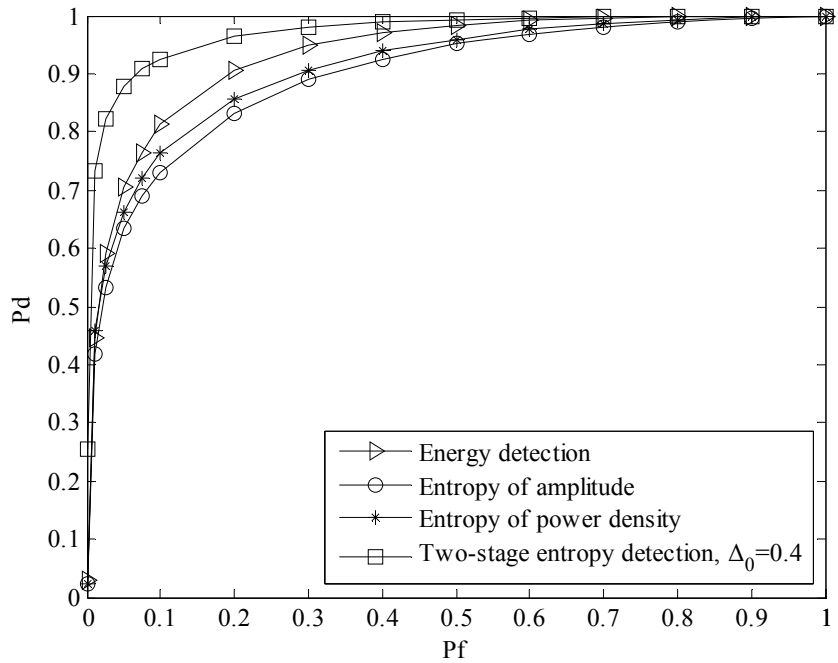

**Fig. 3**. ROC curves comparison of the detectors in Rayleigh fading channel without noise uncertainty when SNR is equal to -10dB

From the simulations above, we can see that the detection performance of the energy detector is better than the one-stage entropy detectors, however, the entropy detectors are robust to the noise uncertainty while the energy detection is very sensitive to the variation of the background noise. The Pd and Pf performance of the energy detector and two-stage entropy detector is compared in Fig. 4 with the power of the background noise varying from -97dbmW to -93dbmW when SNR is fixed at -12dB. In **Fig. 4**, Pf of the energy detection and the two-stage entropy detection is both equal to 0.1 when the noise power is -95dbmW. Pf and Pd of the two-stage entropy detection remain unchanged with the noise power varying, and noise uncertainty can not affect the performance of the entropy-based detectors. On the other hand, the energy detector is very sensitive to the noise uncertainty, and the Pf and Pd become rather unacceptable with the noise uncertainty larger than only 0.5dbmW. As the background noise fluctuates in almost all the practical communication networks, the energy detection with fixed threshold is hardly suitable in practical systems.

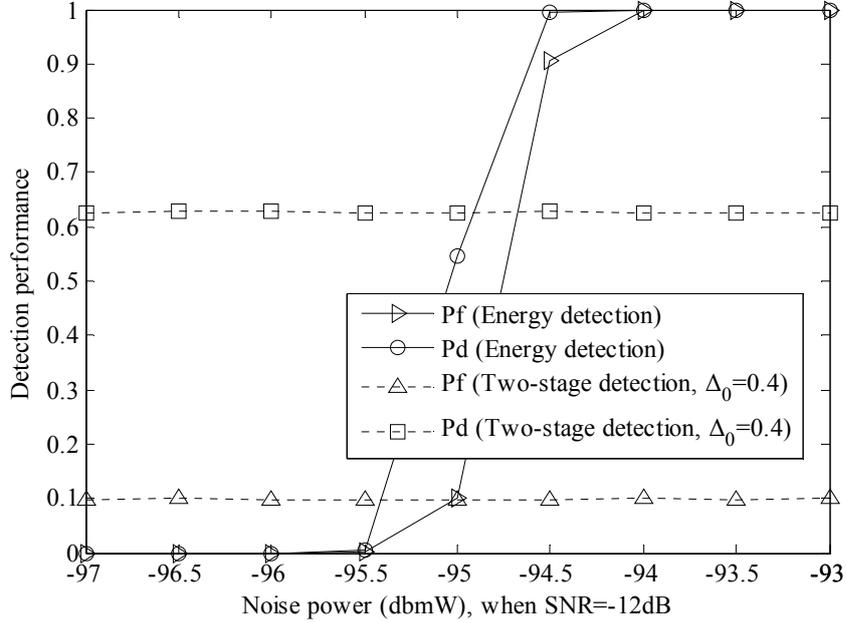

**Fig. 4**. Detection performance with ±2dbmW noise uncertainty when SNR=-12dB

To further compare the computational complex of the one-stage entropy detector and two-stage entropy detector, a computational complex ratio $\Gamma$ is defined as

$$\Gamma = \frac{\text{Computational complex of two-stage entropy detector}}{\text{Computational complex of one-stage entropy detector}}, \quad (25)$$

and ($\Gamma$-1) represents the probability of whether the second stage processing is needed. The computational complex ratio of the two-stage entropy detectors with different $\Delta_0$ against SNR is depicted in **Fig. 5** when Pf is equal to 0.1 and the primary signal exists in the case of $H_1$. It is shown that when SNR is smaller than -7dB and larger than -13dB, the computational complex of the two-stage entropy detection is almost the same as the one-stage entropy detection, and when SNR becomes lower or the primary user is not active, the probability of the second stage processing is bigger and the computational complex of the two-stage entropy detection becomes larger. We can also see that when $\Delta_0$ becomes larger, the computational complex of the two-stage entropy detection increases at certain SNR.

We can also see the computational complex becomes a little smaller when SNR is lower than -13dB. This is because when SNR is extremely low, the received signal is getting close to only WGN received. In this case, the entropy of the spectrum power density of the received signal is prone to be larger than the above threshold $\lambda+\Delta_0$, hence, the computational complex becomes a little smaller. On the other hand, the comutational complex radio $\Gamma$ will not reach 1 even when only WGN is received, because Pf is set to 0.1, and that means even only WGN is received, the probability when detected entropy is smaller than $\lambda+\Delta_0$ in the first stage detection is larger than 0.1.

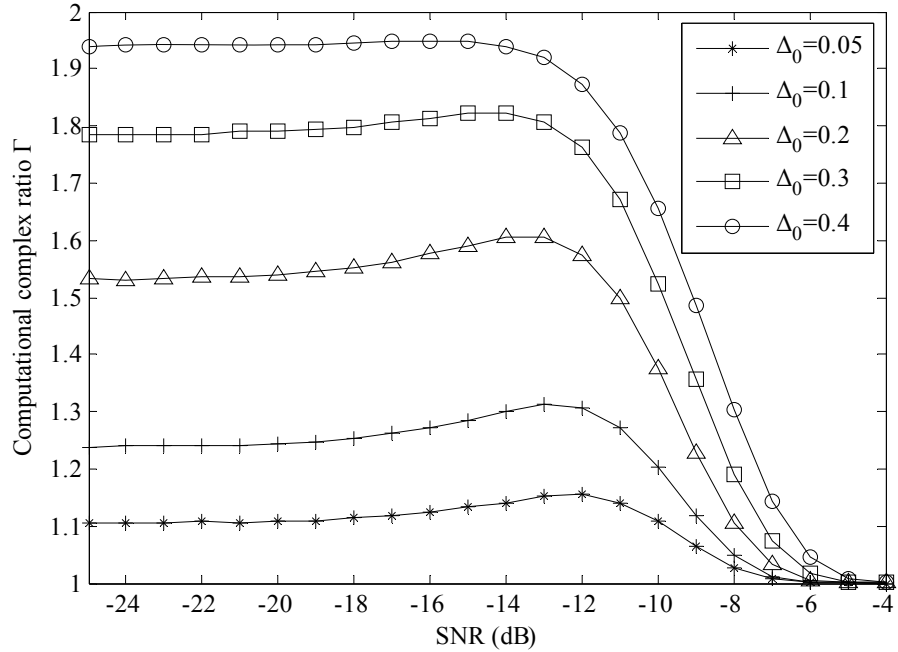

**Fig. 5**. Computational complex ratio Γ against SNR with different $\Delta_0$

From **Fig. 5** we can see that when SNR is low (lower than -13dB) and $\Delta_0$ is relatively big (larger than 0.3), the computational complex of the two-stage entropy detection is close to twice of the computational complex of the one-stage entropy detection. It is also explicit that the computational complex of the one-stage entropy detection with 2*N*-point DFT is almost twice of that of one-stage entropy detection with *N*-point DFT. Hence it is necessary to compare the detection performance of the two-stage entropy detection with *N*-point DFT and one-stage entropy detection with 2*N*-point DFT, and it is shown in **Fig. 6**. In the simulation, Pf is set to 0.1. It is shown that the detection performance of two-stage entropy detection with 1024-point DFT is better than that of the one-stage entropy detection with 2048-point DFT when $\Delta_0$ is larger than 0.1, while the computational complex of two-stage entropy detection with 1024-point DFT is lower than or even half of (SNR>-7dB) that of one-stage entropy detection with 2048-point DFT.

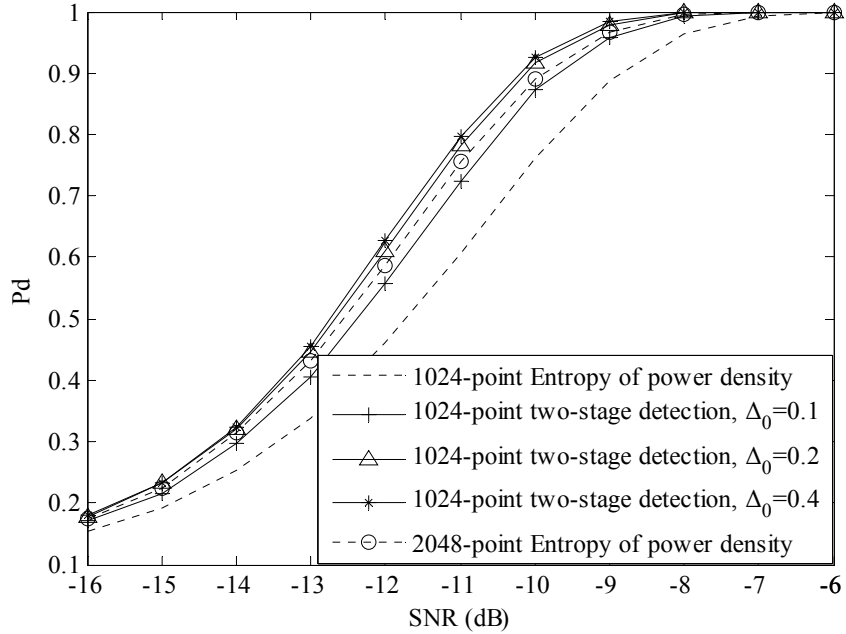

**Fig. 6**. Detection performance comparison of the 1024-point two-stage entropy detection and 2048-point one-stage entropy detection

Then the performance of the proposed cooperative spectrum sensing is evaluated. The Pd performance of some cooperative entropy-based spectrum sensing schemes are compared in **Fig. 7** when Pf=0.1 and SNR is varying from -16dB to -6dB. In the two-stage detection schemes, $\Delta_0$ is both set to 0.3. To make sure that the computational complexity of these entropy-based cooperative spectrum sensing schemes is almost the same, in the cooperative two-stage detection scheme based on entropy, the DFT is 1024-point, while in the cooperative detection with AND, OR, and VOTING rules, the DFT is 2048-point. From **Fig. 5** we can see that the computational complexity of two-stage 1024-point cooperative spectrum sensing scheme is much smaller than that of one-stage 2048-point cooperative spectrum sensing scheme, especially when SNR is relatively large; on the other hand, in the two-stage cooperative spectrum sensing scheme, the decision sent to the common receiver by each secondary user is two bits, while in the one-stage only one-bit decision is sent to the common receiver. Therefore, considering the above analysis, the computational complexity of the proposed two-stage cooperative spectrum sensing with 1024-point DFT is a little smaller than that of the one-stage cooperative spectrum sensing scheme with 2048-point DFT. The performance of two-stage entropy-based detection and one-stage entropy-based detection with only one secondary user and 1024-point DFT is also analyzed.

From the simulation results in **Fig. 7**, we can see that the Pd performance of the proposed cooperative two-stage entropy-based detection scheme with 1024-point DFT is much better than the three traditional cooperative entropy-based detection schemes (AND, OR, and VOTING rules).

The ROC performance of these cooperative entropy-based detectors is also analyzed when SNR is equal to -12dB, and the ROC curves are depicted in **Fig. 8**. The parameters of these detectors are the same as those set in Pd in **Fig. 7** performance analyzed, and it is sure that the computational complex of the cooperative two-stage entropy-based detection scheme with 1024-point DFT and the traditional cooperative one-stage entropy-based detection schemes with 2048-point DFT (AND, OR, and VOTING rules) are almost the same. From the simulation results in **Fig. 8**, we can see that the ROC performance of cooperative two-stage entropy-based detection scheme is much better than that of the other detectors.

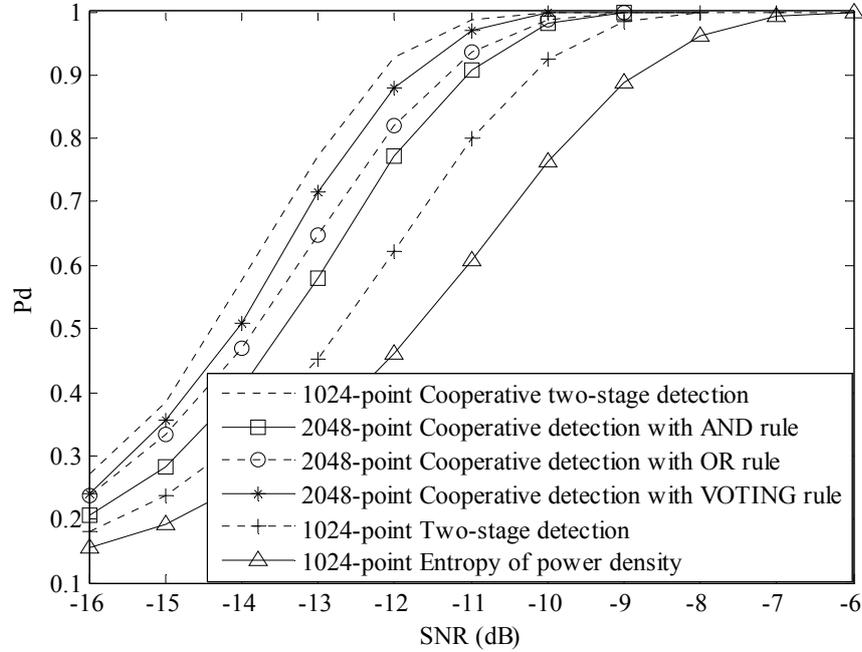

**Fig. 7**. Detection Performance against SNR comparison of the entropy-based cooperative detectors in Rayleigh fading channel without noise uncertainty

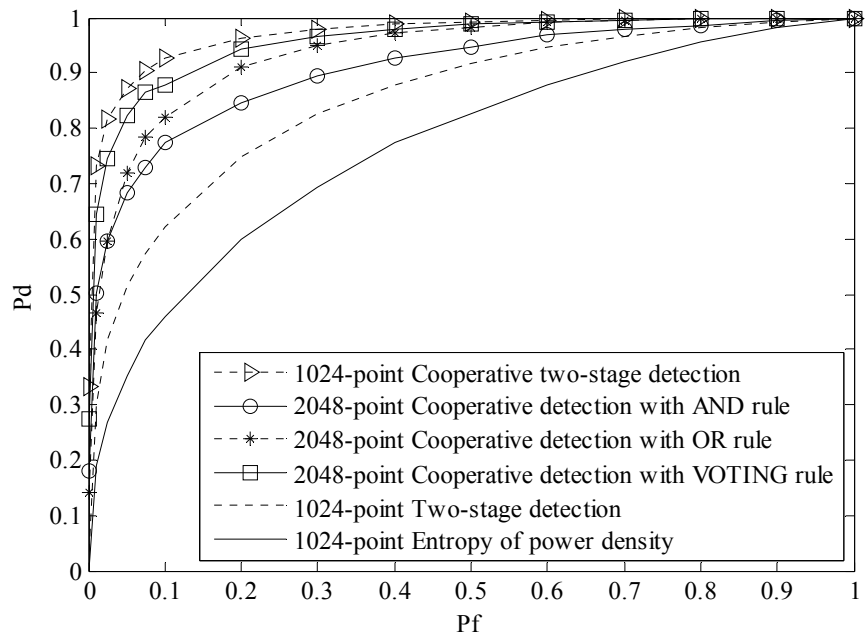

**Fig. 8**. ROC curves comparison of the cooperative entropy-based detectors in Rayleigh fading channel without noise uncertainty when SNR is equal to -12dB

## 6. Conclusion

A two-stage entropy-based robust cooperative spectrum sensing scheme with two-bit decision for cognitive radio is proposed in this letter. First a novel entropy detection based on spectrum power density is designed, and is proved to be robust to the noise uncertainty. The detection performance of the novel entropy detection is shown to be better than the previous entropy detection with lower computational complex. To further improve the reliability of the proposed entropy detection, a two-stage detection scheme is proposed and combined with the proposed entropy detector. Furthermore, to improve the reliability of the detection, a cooperative spectrum sensing scheme with two-bit decision getting from results of the two-stage detection is proposed. It is also shown that the performance of the proposed two-stage cooperative spectrum sensing scheme is much better than the traditional cooperative spectrum sensing schemes with twice of DFT points.

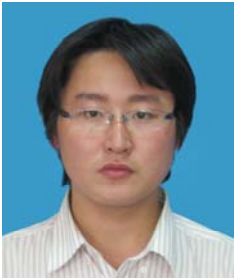

**Nan Zhao** was born in Dalian, China, in 1982. He received the B.S. degree in electronics and information engineering in 2005, the M.E. degree in signal and information processing in 2007, and the Ph. D. degree in Information and communication engineering in 2011 from Harbin Institute of Technology, Harbin, China. He is currently a lecturer at School of Information and Telecommunication Engineering, Dalian University of Technology. His research interests are multiuser detection and power control in CDMA, spectrum sensing in cognitive radio, chaotic theory, and ant colony optimization.